# Unravelling the effect of SrTiO$_3$ antiferrodistortive phase transition on the magnetic properties of La$_{0.7}$Sr$_{0.3}$MnO$_3$ thin films

D. A. Mota,[1] Y. Romaguera Barcelay,[1] A. M. R. Senos,[2] C. M. Fernandes,[2] P. B. Tavares,[3] I. T. Gomes,[4] P. Sá,[4] L. Fernandes,[3] B. G. Almeida,[4] F. Figueiras,[5] P. Mirzadeh Vaghefi,[5] V. S. Amaral,[5] A. Almeida,[1] J. Pérez de la Cruz,[1] and J. Agostinho Moreira[1,*]

[1]IFIMUP and IN-Institute of Nanoscience and Nanotechnology, Departamento de Física e Astronomia da Faculdade de Ciências da Universidade do Porto. Rua do Campo Alegre, 687, 4169-007 Porto, Portugal

[2]Department of Materials and Ceramics Engineering, CICECO, University of Aveiro, 3810-193 Aveiro, Portugal

[3]Centro de Química – Vila Real. Universidade de Trás-os-Montes e Alto Douro. Apartado 1013, 5001-801. Vila Real. Portugal

[4]Centro de Física. Universidade do Minho, P-4710-057 Braga, Portugal

[5]Department of Physics & CICECO, University of Aveiro, 3810-193 Aveiro, Portugal

* Corresponding author: jamoreir@fc.up.pt

**Abstract**

Epitaxial La$_{0.7}$Sr$_{0.3}$MnO$_3$ (LSMO) thin films, with different thickness ranging from 20 nm up to 330 nm, were deposited on (100)-oriented strontium titanate (STO) substrates by pulsed laser deposition, and their structure and morphology characterized at room temperature. Magnetic and electric transport properties of the as-processed thin films reveal an abnormal behavior in the temperature dependent magnetization M(T) below the antiferrodistortive STO phase transition (T$_{STO}$) and also an anomaly in the magnetoresistance and electrical resistivity close to the same temperature. Up to 100 nm LSMO thin films, an in-excess magnetization and pronounced changes in the coercivity are evidenced, achieved through the interface-mediated magnetoelastic coupling with antiferrodistortive domain wall movement occurring below T$_{STO}$. Contrarily, for thicker LSMO thin films, above 100 nm, an in-defect magnetization is observed. This reversed behavior can be understood within the emergence in the upper layer of the film, observed by high resolution transmission electron microscopy, of a branched structure needed to relax elastic energy stored in the film which leads to randomly oriented magnetic domain reconstructions. For enough high-applied magnetic fields, as thermodynamic equilibrium is reached, a fully suppression of the anomalous magnetization occurs, wherein the temperature dependence of the magnetization starts to follow the expected Brillouin behavior.

## I. Introduction

Rare-earth manganite compounds exhibit a strong coupling between their electronic, spin and structural degrees of freedom.[1,2] Due to this coupling, these materials show a rich variety of effects when a change in one of the degrees of freedom induces a response in another.[2-4] An interesting case is found in compounds revealing an interplay between the lattice and the magnetic response, through either spin-phonon or spin-lattice coupling.[5-7] In such materials, the super-exchange interactions, determining the magnetic and electric properties, are found to be strongly dependent on the bond-length of the oxygen $MnO_6$ octahedron and their rotations.[8,9] Deformations (static or dynamic) of these polyhedron modify the spin arrangement and electron transfer underlying the super-exchange interactions. Due to the high sensitivity of the magnetism on lattice coupling, colossal magnetoresistance $La_{0.7}Sr_{0.3}MnO_3$ (LSMO) thin films are good candidates to probe the effects associated with structural changes on magnetotransport and magnetic properties. In $La_{1-x}Sr_xMnO_3$ thin films it is known that lattice mismatch between the manganite and the substrate induces strain in the films, changing both Mn-O bond lengths and Mn-O-Mn bond angles inside the unit cell. This feature turns their structural, transport and magnetic properties strongly dependent on the substrate characteristics.

Epitaxial thin films are found to be exceptional systems to study the interplay between magnetism and crystal lattice. As thin films can be subjected to strong elastic strains due to structural and thermal mismatch with respect to substrates, their intrinsic physical properties can be significantly modified. Strontium titanate (STO) is a cubic perovskite material at room temperature, widely used as substrate for epitaxial growth of manganite films, inducing biaxial strain, either compressive or tensile, depending on the growing thin film. STO undergoes a structural phase transition into an antiferrodistortive phase at $T_{STO}$=105 K, with tetragonal symmetry.[10] This phase transition is driven by an unstable lattice mode of the Brillouin zone border, associated with the rotation of the $TiO_6$ octahedron around a former cubic axis, producing an expansion of the unit cell in the tetragonal *c*-axis, and a contraction in the two other perpendicular directions.[10]

Extensive studies on $La_{1-x}Sr_xMnO_3$ epitaxial thin films, processed onto oriented STO substrates were published for samples having x = 0.33 to 0.47, and the effect of the STO antiferrodistortive phase transition on the magnetic and transport properties has been subject of several reports.[11-14] However, no unique interpretation of the experimental results have, so far, been reached. In fact, according to Segal *et al*,[11] the electrical resistivity cusp and magnetization dip observed in the vicinity of the STO antiferrodistortive phase transition on $La_{0.53}Sr_{0.47}MnO_3$ (LSMO-47) ultra-thin films (11 unit cells) deposited on (100)-STO substrates, are the result of an evanescent cross-interface coupling between the charge carriers in the film and the soft phonon in the STO, mediated through linked oxygen octahedral motions. The motions of the $TiO_6$ octahedra couple to the $MnO_6$ ones, inducing both static and dynamic changes in their configuration.[11] According to these authors, the static effect appears only below $T_{STO}$ and does not explain the anomalous temperature behavior of both resistivity and magnetization.[11] In order to account for these anomalies, a dynamical coupling was assumed.[11] Near $T_{STO}$, the correlation length in STO diverges and the oxygen motions in the interfacial LSMO-47 layer become correlated with the resistivity and magnetization anomalies observed in the film. So, the soft phonons extend into the LSMO-47 layer and couple with magnon modes. However, the penetration depth of these soft phonons was estimated to be just a few (2-3) atomic layers of the film; as such, this effect should not be observed in thicker films.[11]

More recently, Pesquera *et al*[12] reported an experimental study of the dc and ac magnetic properties in $La_{0.67}Sr_{0.33}MnO_3$ thin films (26 and 220 nm thickness) prepared onto (100)-oriented STO substrates. In order to explain the anomalous magnetic responses at low ac-magnetic field and at high dc-field, even in thicker films, magnetic domain pattern reconstruction and creation of regions within the magnetically soft LSMO with enhanced magnetic anisotropy were proposed as two distinct mechanisms, that are triggered by the STO phase transtion.[12] These authors refute the dynamical coupling with the soft phonon assumed in an earlier work,[11] as a similar anomalous temperature behavior of the magnetization is also observed in the thicker films.[12] It is interesting to stress that the LSMO films used in both works have different compositions. This might not enable a straight comparison between the results obtained, since their physical properties may be different, as it can be suggested from the (%Sr, T) phase diagram.[13]

Thin films are under significant strains due to the mismatch between the cell dimensions of both film and substrate. Moreover, the distortions provided by film-substrate mechanical interactions can drastically modify the physical properties of the films when compared with the bulk material. This is particularly important for rare-earth manganites, whose magnetic properties can be deeply changed by external parameters. The effect of the pinning to the substrate on the magnetic properties and its strength with the film thickness, has not been yet explored in LSMO thin films on STO substrates, under weak external magnetic fields. In this work, we present a detailed study regarding the effect of the emergence of antiferrodistortive domains, occurring below ~105 K in the (100)-oriented STO substrate, on the magnetic and magnetoresistive properties of LSMO thin films as a function of thickness, taking values between 20 and 330 nm. After performing a detailed structural analysis of the as-processed LSMO films, the study of the temperature dependence of magnetization as a function of dc magnetic field as well as of magnetic inversion curves for different film thicknesses will be undertaken in order to sort out any changes in the magnetic behavior of LSMO films, which might be induced by antiferrodistortive ordering occurring in the STO substrate.

## II. Experimental

High quality ceramic LSMO targets were prepared by the solid-state method, according to Ref. 14. Epitaxial LSMO thin films with thicknesses between 20 and 330 nm were grown on single crystalline oriented (100)-STO substrates by pulsed laser deposition. The depositions were carried out with a KrF excimer laser (wavelength λ = 248 nm), at a fluence of 1.5 J/cm$^2$, with a 3 Hz repetition rate and pulse duration of 25 ns. The target-to-substrate distance was kept constant at 5.5 cm during the depositions, which occurred in a pure oxygen atmosphere with 0.8 mbar pressure and with a substrate temperature of 700 °C. To prevent under-oxidation of the films, after the deposition, they were cooled down to room temperature at 15 °C/min in a pure oxygen atmosphere, at atmospheric pressure.

Atomic Force Microscopy (AFM) measurements were performed in a Nanoscope IVa multimode system in the tapping mode using Si tips. Scanning Electron Microscopy (SEM) measurements were performed in a FEI Quanta 400 FEG ESEM scanning electron microscope operated at 20 kV and with a resolution of 1.5 nm for both surface morphology and film thickness characterization. High Resolution X-Ray Diffraction[15] measurements were performed in a XPERT-PRO Diffractometer, using the main X-Ray line: Cu$_{k\alpha 1}$ = 1.5405980 Å through a 4xGe220 Asym. monochromator.

The dc magnetic properties were measured with a Quantum Design MPMS SQUID magnetometer, with a reciprocating sample option and with a resolution up to $10^{-7}$ emu. Electrical resistivity and magnetoresistance measurements were performed in the temperature ranges 25-300 K and 80-300 K, respectively. In both cases, a standard four probe method was used. A magnetic field of 1 T applied parallel to the electrical current was used for the magnetoresistance measurements.

Cross-sectional samples were thinned by mechanical polishing to about 100 μm and then ion milled using a JEOL EM-09100 IS ion slicer until electron transparency. HRTEM studies were performed by using a FEI Technai electron microscope operated at 200 kV, owing a 0.17 nm spatial resolution.

## III. Results and Discussion

The surface analysis of the 60 nm LSMO thin film, through AFM and SEM, respectively, evidences a uniform topography, with a root mean square roughness of approximately 1 nm, which is of the same order of magnitude even for scans over 2×2 μm$^2$, revealing an almost layer-by-layer growth with some islands nucleation at the surface and showed also that grain-size takes values between 30 to 50 nm, defining sharp squared crystallites with common alignment of edges. Similar results were ascertained for all the as-processed LSMO thin films.

Figure 1 shows a representative X-ray diffraction spectrum obtained for the case of the 210 nm LSMO film. It evidences that the film can be indexed to the rhomboedral R-*3c* (167) space group, having a clear preferential growth following the {100} orientation of the cubic P$m$-$3m$ (221) SrTiO$_3$ substrate.

For the 60 nm LSMO thin film, the HR-XRD results point to approximately 0.62% compressive deformation of (110)$_R$ planes family relative to the conventional bulk form ($a$ = 5.472 Å, $\alpha$ = 60.354°)[16] estimated from the pseudomorphic cell parameter <$a$> ~ 5.45(0) Å. In the representative 2D asymmetric pole figure (figure 2(a)) performed at 2$\theta$ = 32.60° is possible to observe the 4 typical reflections assigned to planes (101)$_R$ and (211)$_R$ of LSMO film neighboring the Bragg reflection assigned to the plane (110)$_C$ of STO substrate. A similar $\psi$($d_{hkl}$) is observed through the superposition of reflections, whereas the slight spread in the in-plane $\varphi$ angle can be interpreted as a minor rotation between these planes. The $\Omega$ vs. 2$\theta$ map, shown in figure 2(b), is centered at 2$\theta$ = 46.50° and $\Omega$ = 23.25°. It involves the STO (200)$_C$ peak and a single side peak of the (220)$_R$ LSMO thin film, which closely follows the 2$\theta$ and $\Omega$ contours from the cubic substrate. This feature evidences both the high degree of epitaxy and coherent growth of this film.

For the 210 nm LSMO thin film the HR-XRD results for $d$(110)$_R$ denote approximately 0.44% compressive deformation relative to the conventional bulk form, with a pseudomorphic cell parameter <$a$> ~ 5.46(0) Å. The respective 2D asymmetric pole figure performed at 2$\theta$ = 32.60° (figure 3(a)) suggests some relaxation of (101)$_R$ and (211)$_R$ planes of LSMO film relative to the substrate (110)$_C$ planes denoted by the four peaks elongation in $\psi$ angle. The $\Omega$ vs. 2$\theta$ map shown in figure 3(b) is centered at 2$\theta$ = 46.50° and $\Omega$ = 23.25° and covers the STO (200)$_C$ peak and the (220)$_R$ peak of the LSMO film. The strong spread $\Delta\Omega$ (~2°) of this latter peak, diverging from the substrate peak shape, evidences the existence of a significant tension, which yields a partially bending of the film planes. Moreover, the elongation of the peak in $\Delta\theta$ (~0.5°) can be associated with $d$(220)$_R$ plane relaxation across the film growth direction. This relaxation may not be homogeneous as observed by Ranno *et al.*,[17] and further confirmed by our HRTEM analysis.

Figure 4(a) and (b) show the magnetization of LSMO films as a function of the applied magnetic field parallel to the film plane, recorded at two different fixed temperatures, above and below T$_{STO}$, for 60 nm and 210 nm film thickness, respectively. These results are illustrative examples of the magnetic response of LSMO films for the various thickness films studied. The magnetization obtained in these conditions is one order of magnitude larger than the magnetization measured in out-of-plane magnetic field conditions (not shown), giving experimental evidence for a magnetization mostly in the film plane. Therefore, we will mainly focus our attention on the results of the magnetization measurements with the applied magnetic field in this direction. The LSMO films, whose results are presented in Figure 4, show typical ferromagnetic hysteresis loops and one remarkable issue is the large variation of the magnetic coercivity of the 60 nm-thick film between 120 K and 60 K.

In order to get a deeper insight regarding the variation of the coercive field, we have plotted it as a function of film thickness as it is shown in figure 5. It can be observed that for lower thicknesses (20 and 60 nm) the coercive field measured at 60 K is much larger than the one measured at 120 K. However, the difference between the coercive fields measured at 60 K and 120 K decreases as the film thickness increases. Moreover, while the coercive field measured at 120 K seems to be weakly thickness dependent, the coercive field measured at 60 K is highly thickness dependent.

These results point out for two outcomes. The first is the existence of a "critical thickness" close to 100 nm above which the coercivity difference at the two temperatures is smaller. The second outcome is the strong effect of the STO phase transition on the magnetic properties of the thinner films. In fact, the films with thickness below 100 nm are the most sensitive to structural changes of the substrate, and clearly present a remarkable change of coercive field upon cooling through $T_{STO}$. The change in the coercive field results in a stabilization of a distinct magnetic domain configurations at remanence, or when small magnetic fields are applied to the samples. The magnetic properties of such thin films are tailored by the effect of the antiferrodistortive phase transition of the STO substrate which controls, in a very specific way, both the coercive field and magnetization. So, the comparison of M(H) hysteresis loops obtained at 60 K and 120 K must be hold carefully, because they have been registered in different physical conditions regarding the STO state. Looking at the experimental results, a close relation between the structural units of both STO and LSMO film is clearly evident and the structural coupling between STO and film ensured through the STO/LSMO interface is the driving force of the magnetic properties of the LSMO films. In the thinner films, the increase of the coercive field observed at 60 K, relatively to the value at 120 K, is a consequence of the strong pinning of the magnetic domains to the substrate, which drastically changes its domain pattern between these two temperatures. Moreover, the coupling between both structures may influence the dc magnetization on crossing the $T_{STO}$ phase transition. Actually, the impact of the substrate-induced changes is smaller in thicker films due to strain relaxation and possibly to the presence of different layers in thicker films.

Figure 6 shows the HRTEM cross-section of a 330 nm thick film. The existence of two different layers can be distinctly observed. A first one with a thickness close to 100 nm, wherein the film grows epitaxially and coherently to the STO substrate (inset B1), and a second one that is not directly linked to the substrate and where the film shows a branched structure (inset A) associated with relaxation processes, also observed in XRD analysis. The brighter reflections on the electron diffraction pattern (inset B2 of figure 6) are attributed to STO substrate and the others belong to LSMO film. From the analysis of these reflections, an epitaxial growth LSMO (02-2)∥STO (100) relation could be inferred.

The effect of the antiferrodistortive phase transition occurring in STO substrate changes other physical properties of the films. The electric resistivity and magnetoresistance curves present an anomaly at $T_{STO}$ = 105 K, as shown in figure 7, which demonstrate that they are sensitive to domain wall scattering and corroborates that the anomalous behavior of the temperature dependence of the coercivity is a genuine effect of the stabilization of the antiferrodistortive phase of the STO substrate below $T_{STO}$=105 K.

Is it well established that the magnetization, measured as a function of temperature, is very sensitive to domain patterns, and more specifically, to magnetic domain reconfiguration taking place at the antiferrodistortive phase transition occurring at $T_{STO}$. This is particularly important as the pinning between the magnetic domains with the substrate, which induces strain on the film, can modify the film coercivity.[18]

Figures 8(a) and 8(b) show the dc magnetization for the 60 nm and 210 nm LSMO film, respectively, measured after field cooling the sample with an applied magnetic field of 100 Oe, parallel to the film plane. The temperature dependence of the dc magnetization, recorded above 110 K, follows a smooth behavior as shown by the auxiliary red dots in figures 8(a) and 8(b). The film exhibits a ferromagnetic-like behavior, with a Curie temperature $T_C$ = 344K, which is in good agreement with earlier published results.[12] Below $T_{STO}$ = 105 K, the magnetization exhibits two distinct anomalous behavior, starting to deviate from the red smooth dots, depending on the film thickness.

We observed that for LSMO thin films with thicknesses up to close 100 nm have a similar magnetic response to the one shown in figure 8(a). In the 60 nm thick, the magnetic anomaly observed in figure 8(a) below $T_{STO}$ is interpreted, as we observed before, by taking into account the epitaxial and coherent film growth which results in the change of the coercive field and thus different magnetic domain configurations at remanence, or when small magnetic fields are applied to the samples at the $T_{STO}$.

LSMO thin films with thicknesses above 100 nm have a similar magnetic response to the one shown in figure 8(b). As we have concluded before and observed by the HRTEM analysis, the impact of the substrate-induced changes are smaller in thicker films which is associated with strain relaxation and to the presence of two different "layers". The emergence of an in-defect magnetization could be understood by bringing into discussion magnetic domain reconstructions of the soft LSMO matrix coming from the upper relaxed "layer" among other plastic defects observed in HRTEM analysis. It is worth noting that this reconstruction of magnetic domains was also proposed by Pesquera *et al*.[12]

It is important to stress that the applied magnetic field used in the M(T) measurements has a magnitude of 100 Oe which is below the saturation field (see figure 4), and thus is not sufficient high to reach fully reorientation of the spins. Figure 9 shows the M(T) curves for the 210 nm thick film for several applied magnetic fields. Whereas the in-defect magnetization continuously decreases as the magnetic field rises from 100 to 400 Oe, already above a magnetic field of 400 Oe a thermodynamic state is reached where the magnetization follows a typical Brillouin behavior. Increasing the magnetic field above the saturation value leads to the one direction orientation of all magnetic domains and consequently the anomalous magnetic behavior observed below $T_{STO}$ is suppressed.

IV. Conclusions

In summary, this work evidences that antiferrodistortive ordering emerging in STO below 105 K, induces significant changes on the magnetic microstructure of the LSMO thin films. This is especially remarkable in thin films, thinner than 100 nm, due to the lack (or rather smaller) density of plastic defects, among other, associated with the strain relaxation and due to the strong pinning of the magnetic domains of the LSMO film to the substrate domain pattern. These magnetic changes are most evident in the field cooling magnetization and in the coercivity, which largely increases when lowering temperature below $T_{STO}$. Contrarily, the in-defect magnetization observed for thicker films can be understood by the formation of randomly oriented magnetic domain reconstructions associated with film relaxations confirmed by both XRD and HRTEM. A fully suppression of the anomalous magnetization occurs for high enough applied magnetic fields, wherein a thermodynamic equilibrium is reached. The temperature dependence of the magnetization then follows the usual Brillouin behavior.

**Acknowledgments**

This work was supported by the Fundação para a Ciência e Tecnologia and COMPETE/QREN/EU, through the project PTDC/CTM/099415/2008. The authors are very grateful to Maria João Pereira and Maria Rósario Soares from CICECO, University of Aveiro, for the HR-XRD measurements and discussion of the results. F. Figueiras acknowledges FCT grant SFRH/BPD/80663/2011. The authors also acknowledge Projeto Norte-070124-FEDER-000070 and Prof. J. Fontcuberta for fruitful discussions.

**Figure Captions**

**Figure 1.** X-ray diffraction pattern obtained for the 210 nm LSMO film sample, displaying the respective indexations of the SrTiO$_3$ (STO) substrate to the Cubic *Pm-3m* (211) symmetry group, and of the LSMO thin film to the Rhomboedric *R-3c* (167) pseudomorphic phase.

**Figure 2**. 2D asymmetric pole figure at 2θ = 32.6° of the 60 nm LSMO thin film (a) and map centered at 2θ= 46.50° and Ω = 23.25° (b).

**Figure 3.** 2D asymmetric pole figure at 2θ = 32.6° of the 210 nm LSMO film (a) and map centered at 2θ = 46.50° and Ω = 23.25° (b).

**Figure 4.** Field dependence of magnetization measured at T = 60 K and 120 K along the [100] in-plane direction for the 60 nm (a) and 210 nm (b) LSMO thin film.

**Figure 5.** Thickness dependence of the coercive field measured at 60 and 120 K.

**Figure 6.** HRTEM picture of a cross-sectional sample of 330 nm LSMO thin film deposited on STO substrate. A) Top layer showing a branched structure caused by the relaxation of the film; B1) High magnification TEM image, and B2) electron diffraction pattern of the interface STO/LSMO.

**Figure 7.** Temperature dependence of the electric resistivity of the 60 nm LSMO thin film. The inset shows a zoom image of the electrical resistivity and magnetoresistance close to the T$_{STO}$ of the 60 nm LSMO thin film measured under a magnetic field of 1 T applied parallel to the electrical current.

**Figure 8.** Temperature dependence of the 60 nm (a) and 210 nm (b) LSMO thin film magnetization, measured in field-cooling conditions with 100 Oe applied magnetic field in-plane.

**Figure 9.** Temperature dependence of the magnetization of the 210 nm LSMO thin film measured in field-cooling conditions with different applied in-plane magnetic fields.

**Figures**

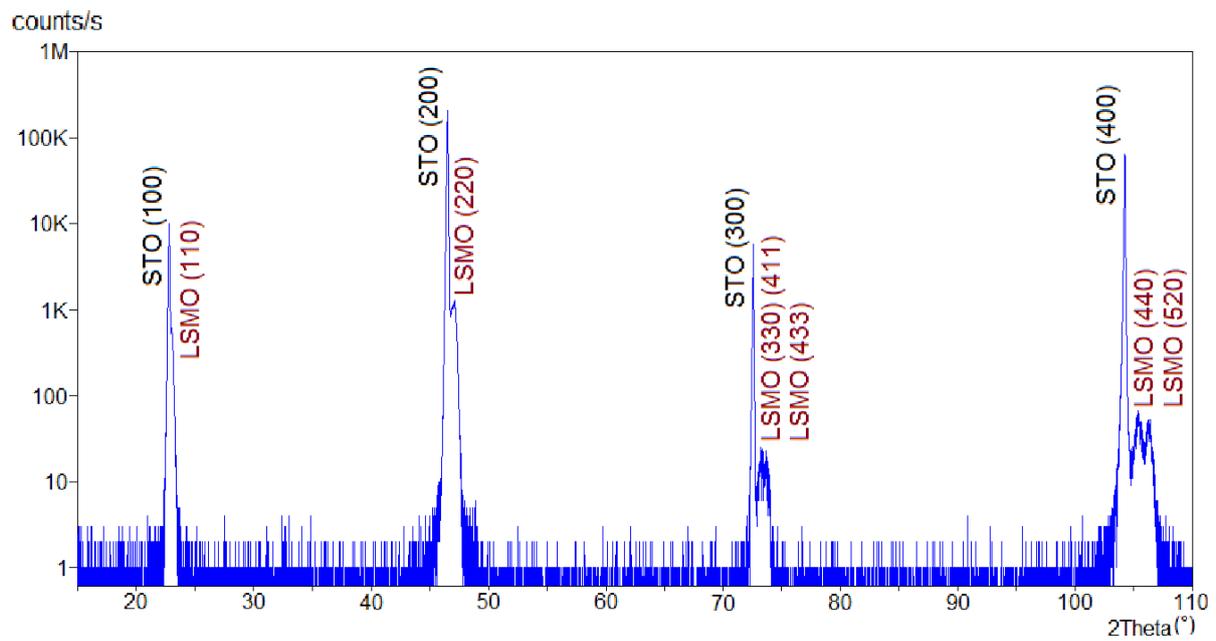

Figure 1.

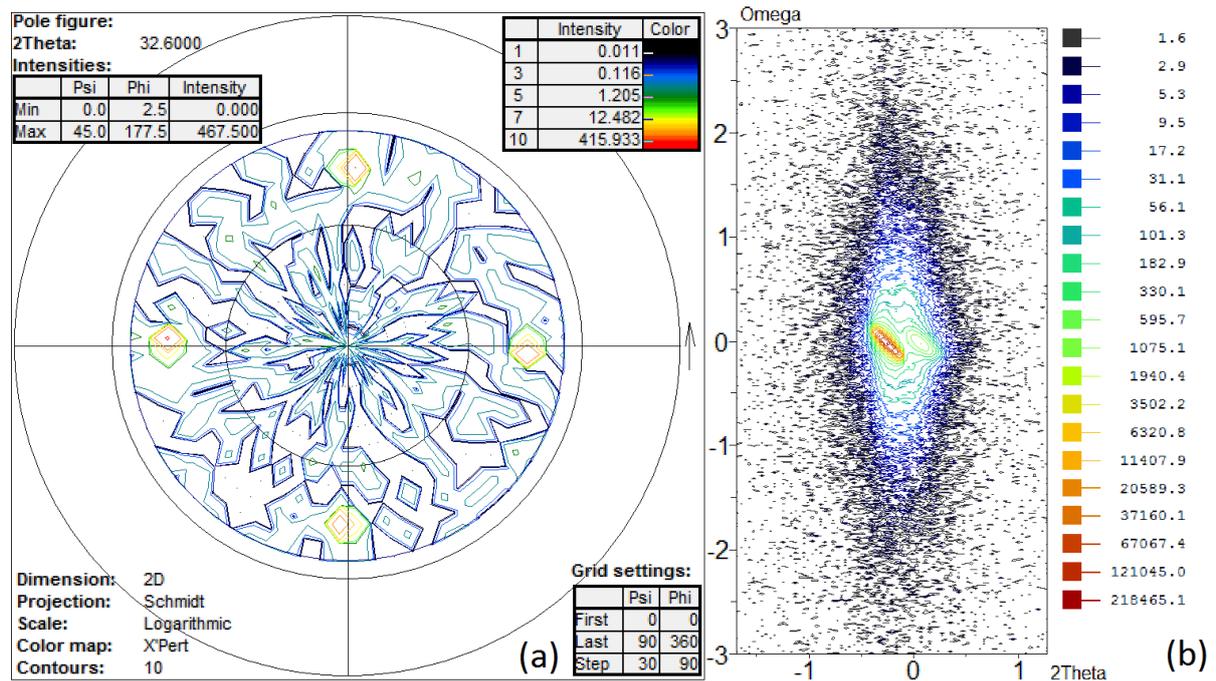

Figure 2.

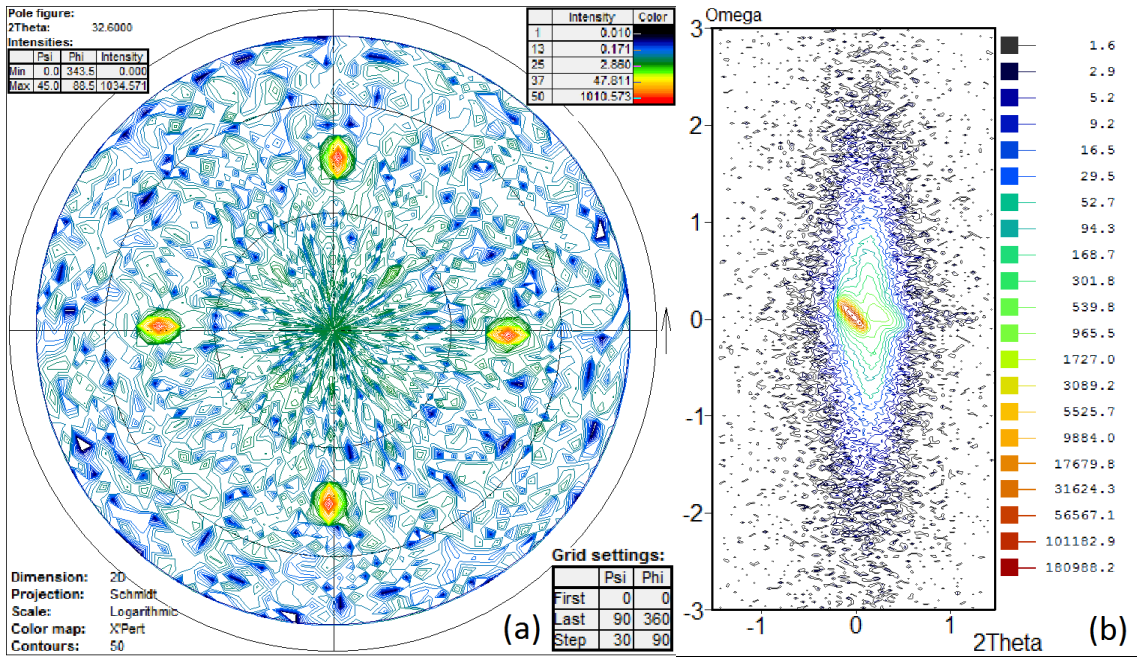

Figure 3.

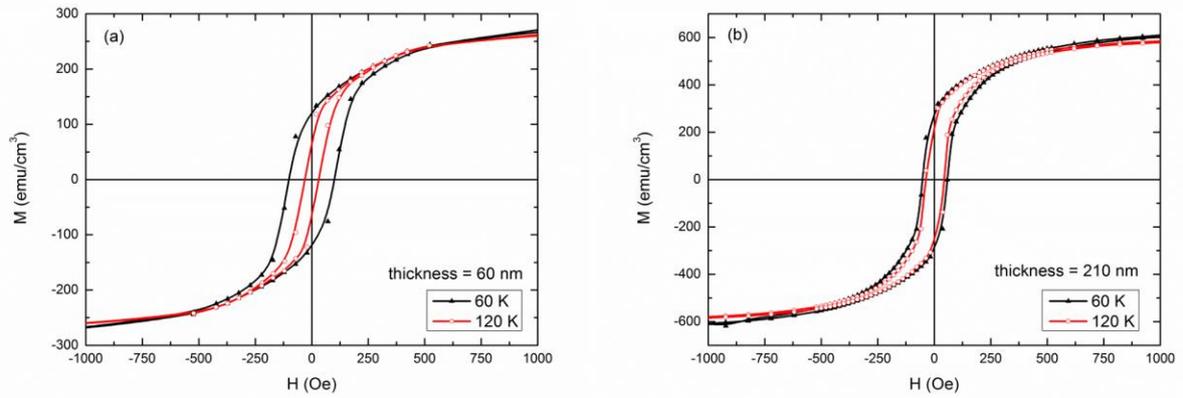

Figure 4.

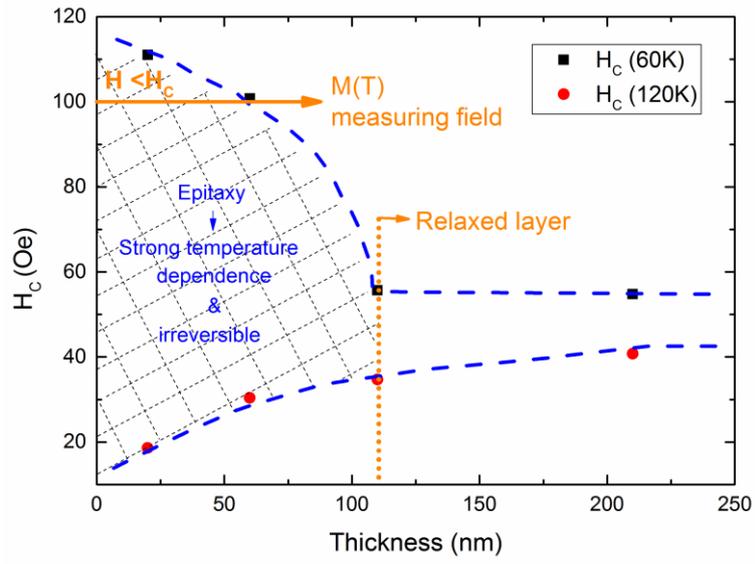

Figure 5.

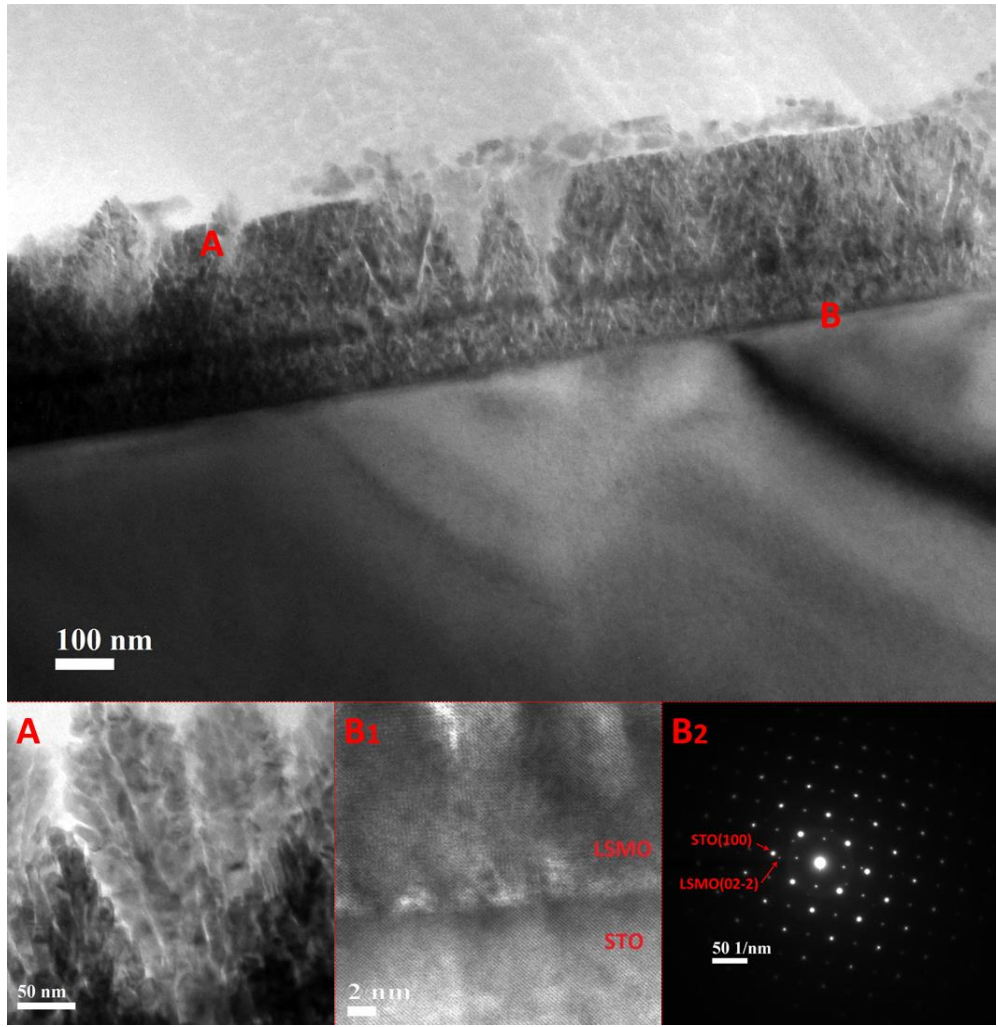

Figure 6.

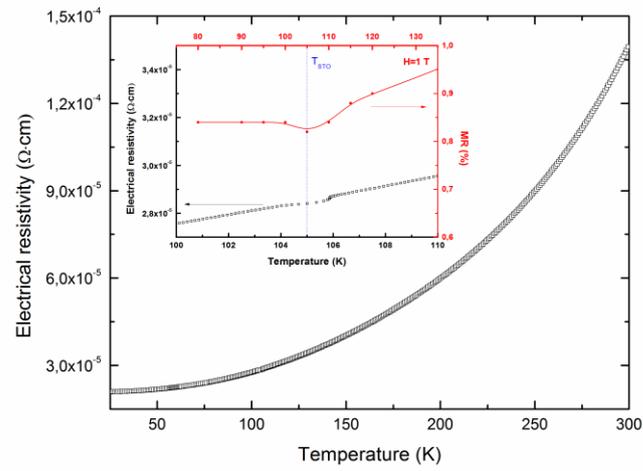

Figure 7.

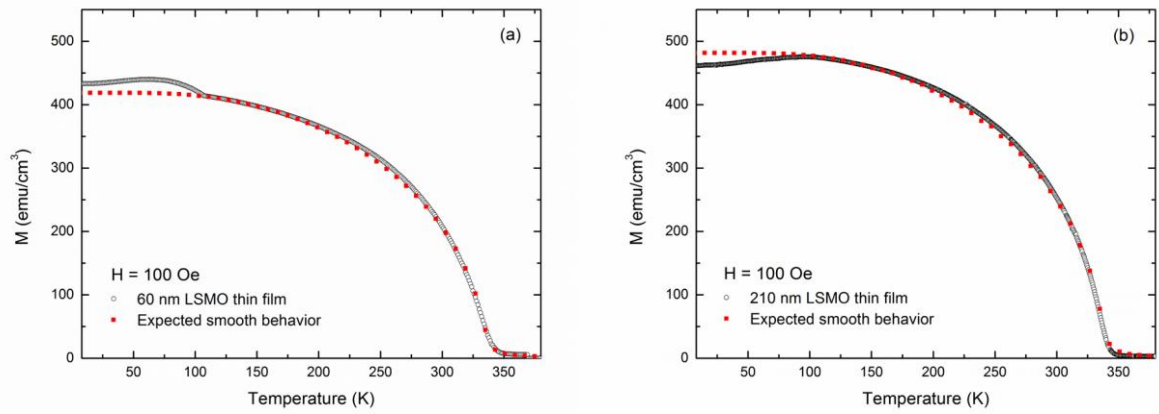

Figure 8.

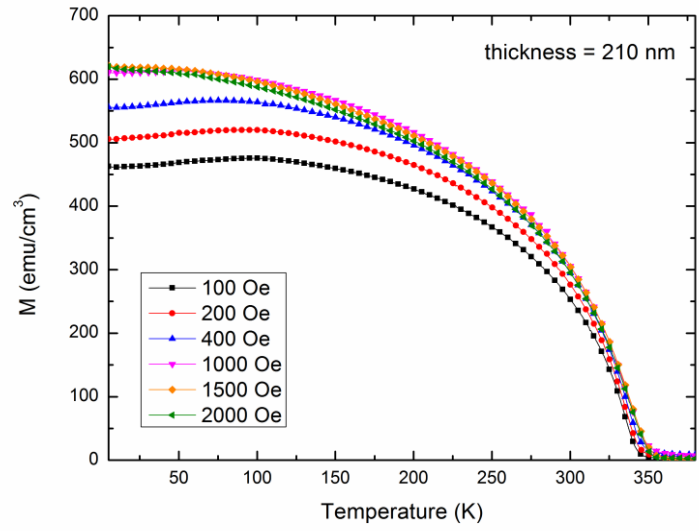

Figure 9.